\documentclass[aps,prd,twocolumn,nofootinbib]{revtex4-1}

\usepackage{graphicx}

\usepackage[colorlinks=true, linkcolor=blue, citecolor=blue, urlcolor=blue]{hyperref}

\begin{document}

\title{The Gross-Llewellyn Smith sum rule up to ${\cal O}(\alpha_s^4)$-order QCD corrections}

\author{Xu-Dong Huang}
\email{hxud@cqu.edu.cn}

\author{Xing-Gang Wu}
\email{wuxg@cqu.edu.cn}

\author{Qing Yu}
\email{yuq@cqu.edu.cn}

\author{Xu-Chang Zheng}
\email{zhengxc@cqu.edu.cn}

\author{Jun Zeng}
\email{zengj@cqu.edu.cn}

\affiliation{Department of Physics, Chongqing University, Chongqing 401331, People's Republic of China}

\date{\today}

\begin{abstract}

In the paper, we analyze the properties of Gross-Llewellyn Smith (GLS) sum rule by using the $\mathcal{O}(\alpha_s^4)$-order QCD corrections with the help of principle of maximum conformality (PMC). By using the PMC single-scale approach, we obtain an accurate renormalization scale-and-scheme independent fixed-order pQCD contribution for GLS sum rule, e.g. $S^{\rm GLS}(Q_0^2=3{\rm GeV}^2)|_{\rm PMC}=2.559^{+0.023}_{-0.024}$, where the error is squared average of those from $\Delta\alpha_s(M_Z)$, the predicted $\mathcal{O}(\alpha_s^5)$-order terms predicted by using the Pad\'{e} approximation approach. After applying the PMC, a more convergent pQCD series has been obtained, and the contributions from the unknown higher-order terms are highly suppressed. In combination with the nonperturbative high-twist contribution, our final prediction of GLS sum rule agrees well with the experimental data given by the CCFR collaboration.

\end{abstract}

\maketitle

The Gross Llewellyn-Smith (GLS) sum rule indicates that the isospin singlet structure function $xF_3(x,Q^2)$ satisfies an unsubtracted dispersion relation~\cite{Gross:1969jf}, and in the quark-parton model, it is equal to the number of valence quarks inside a nucleon. On the experimental side, the GLS sum rule has been firstly measured at $Q^2=Q^2_0=3 {\rm GeV^2}$~\cite{Leung:1992yx}, which gives $S^{\rm GLS}(Q_0^2)=2.50\pm0.018 ({\rm stat.})\pm0.078({\rm syst.})$, where the first and second errors are statistical and systematic errors, respectively. Lately, Ref.\cite{Kim:1998kia} extracts a set of values for the GLS sum rule at various $Q^2$ values by using the neutrino deep-inelastic scattering data at low $Q^2$-region. On the theoretical side, the GLS sum rule for the polarized deep-inelastic electron scattering can be written as
\begin{eqnarray}
S^{\rm GLS} &=& \frac{1}{2} \int_0^1 \frac{d x}{x} xF_3(x,Q^2) = 3 \left[ 1- a_{F_3}(Q) \right] + \frac{\Delta^{\rm HT}}{Q^2}, \label{eq:gls}
\end{eqnarray}
where $xF_3(x,Q^2)=xF^{\nu p}_3(x,Q^2)+xF^{\bar{\nu} p}_3(x,Q^2)$, the effective charge $a_{F_3}(Q)$ represents the perturbative contribution to the leading-twist part, and $\Delta^{\rm HT}$ is the power suppressed non-perturbative higher-twist correction~\cite{Fajfer:1985fw, Ross:1993gb, Braun:1986ty, Anselmino:1995re, Dasgupta:1996hh, Kataev:1999bp, Nath:2016phi}. In the present paper, we shall concentrate our effects on obtaining precise pQCD correction to the GLS sum rule, and we adopt three typical higher-twist corrections to do our discussion, $\Delta^{\rm HT}_1\simeq-0.15\pm0.15$ GeV$^2$~\cite{Kim:1998kia}, $\Delta^{\rm HT}_2 \simeq -0.471\pm 0.036$ GeV$^2$~\cite{Ross:1993gb}, and $\Delta^{\rm HT}_3 \simeq-0.293\pm0.142$ GeV$^2$~\cite{Braun:1986ty}. It is important to finish as more perturbative terms as possible so as to achieve a more accurate pQCD prediction. At the present, the pQCD correction for the leading-twist part of the GLS sum rule has been computed up to ${\cal O}(\alpha_s^2)$~\cite{Gorishnii:1985xm, Zijlstra:1993sh}, ${\cal O}(\alpha_s^3)$~\cite{Larin:1991tj, Chyla:1992cg, Hinchliffe:1996hc, Nath:2015sxi}, and ${\cal O}(\alpha_s^4)$~\cite{Baikov:2010je, Baikov:2012zn}, respectively. Those improvements on perturbative calculations provide the people a good chance of achieving accurate GLS sum rule, as is the purpose of the present paper.

Following standard renormalization group invariance, an infinite-order pQCD prediction is independent to the choices of renormalization scheme and scale, which is ensured by mutual cancelation of the scale dependence among different orders. For a fixed-order pQCD prediction, if using the ``guessed" renormalization scale as conventional scale-setting approach does, one will meet the renormalization scale and scheme ambiguities due to the mismatching of the magnitudes of $\alpha_s$ and its coefficients at each perturbative order~\cite{Wu:2013ei, Wu:2014iba, Wu:2019mky}. Different from conventional scale-setting approach, the principle of maximum conformality (PMC)~\cite{Brodsky:2011ta, Brodsky:2012rj, Mojaza:2012mf, Brodsky:2013vpa} has been suggested to eliminate such scale ambiguities. It has been demonstrated that by using the PMC single-scale approach raised in year 2017~\cite{Shen:2017pdu}, the resultant conformal series is independent to any choice of renormalization scale~\cite{Wu:2018cmb}, and the residual scale dependence~\cite{Zheng:2013uja} due to uncalculated higher-order terms shall be highly suppressed, which generally suffers from both $\alpha_s$-power suppression and exponential suppression. Many successful PMC single-scale approach examples have been done in the literature, and in the paper, we shall adopt it to analyze the perturbative part of the GLS sum rule.

The pQCD series of the effective charge $a_{F_3}$ can be written as the following form,
\begin{eqnarray}
a_{F_3}(Q) &=& r_{1,0}a_s + (r_{2,0}+\beta_{0}r_{2,1})a_s^{2}\nonumber\\
&&+(r_{3,0}+\beta_{1}r_{2,1}+ 2\beta_{0}r_{3,1}+ \beta_{0}^{2}r_{3,2})a_s^{3}\nonumber\\
&& +(r_{4,0}+\beta_{2}r_{2,1}+ 2\beta_{1}r_{3,1} + \frac{5}{2}\beta_{1}\beta_{0}r_{3,2} \nonumber\\
&&+3\beta_{0}r_{4,1}+3\beta_{0}^{2}r_{4,2}+\beta_{0}^{3}r_{4,3}) a_s^{4}+\mathcal{O}(a^5_s), \label{cf3ij}
\end{eqnarray}
where $a_s=\alpha_s/4\pi$, $r_{i,0}$ are conformal coefficients, and the non-conformal coefficients $r_{i,j}$ $(j\neq0)$ which are proportional to the $\{\beta_{i=0,1,\cdots}\}$-functions. Those non-conformal terms can be adopted to get exact $\alpha_s$-value with the help of RGE, for convenience, one can put the log-term $\ln\mu^{2}_{r} /Q^2$ out of the coefficients $r_{i,j}$, i.e.
\begin{equation}
r_{i,j}=\sum^j_{k=0}C^k_j{\hat r}_{i-k,j-k}{\rm ln}^k(\mu_r^2/Q^2), \label{rijrelation}
\end{equation}
where $C^k_j=j!/[k!(j-k)!]$ is the combination coefficient. It is noted that ${\hat r}_{i,0}=r_{i,0}$, and ${\hat r}_{i,j}=r_{i,j}|_{\mu_r=Q}$. Those coefficients can be derived from Refs.\cite{Baikov:2008jh, Baikov:2010je, Baikov:2012zn}, and we put them in the Appendix. After applying the standard PMC single-scale approach~\cite{Shen:2017pdu}, all non-conformal $\{\beta_i\}$-terms, which are used to determine the correct $\alpha_s$-value of the process, can be eliminated from the perturbative series, and the effective charge $a_{F_3}(Q)$ thus becomes a scheme-independent conformal series
\begin{eqnarray}
a_{F_3}(Q)|_{\rm PMC} &=& {\hat r}_{1,0}a_s(Q_*) +{\hat r}_{2,0}a_s^{2}(Q_*) +{\hat r}_{3,0}a_s^{3}(Q_*) \nonumber \\
&&+{\hat r}_{4,0}a_s^{4}(Q_*)+\mathcal{O}(a^5_s), \label{f3PMC}
\end{eqnarray}
where $Q_*$ is the PMC scale, which is determined by requiring all non-conformal terms vanish. Practically, one can expand $\ln Q^2_*/Q^2$ as a power series over $a_s(Q)$, i.e.,
\begin{equation}
\ln{Q^2_*}/{Q^2}=T_0+T_1 a_s(Q)+T_2 a^2_s(Q)+ \cdots, \label{scaleQs}
\end{equation}
whose first three coefficients $T_i$ are put in the Appendix. $Q_*$ can be determined up to next-to-next-to-leading log (N$^2$LL) accuracy by using the known $\mathcal{O}(\alpha_s^4)$-level perturbative series of GLS sum rule. Eq.(\ref{scaleQs}) shows that $Q_*$ is exactly free of $\mu_r$ at any fixed order and determines the effective value of $\alpha_s$, which represents the correct momentum flow of the process. In this sense, we can treat $Q_*$ as the ``physical" scale of the process. Together with the $\mu_r$-independent conformal coefficients, the resultant PMC pQCD series exactly avoids the conventional renormalization scale ambiguity, then the precision of GLS sum rule can be greatly improved.

We take the four-loop $\alpha_s$-running behavior with the reference point $\alpha_s(M_Z)=0.1179$~\cite{Zyla:2020zbs} to do the numerical calculation. We adopt $Q^2=Q_0^2=3$ GeV$^2$ as an explicit example to show how the perturbative nature of GLS sum rule can be improved by applying the PMC.

In small energy scale, $\alpha_s$ shall have infrared divergence, and its running behavior derived from RGE may overestimate the pQCD prediction. Several low-energy models have been suggested to deal with the $\alpha_s$-value at small energy scale~\cite{Brodsky:2010ur, Webber:1998um, Shirkov:2012ux, Badalian:2001by, Cornwall:1981zr, Godfrey:1985xj, Shirkov:1997wi}. A comparison of various low-energy models has been done in Ref.\cite{Zhang:2014qqa}, which indicates that the massive perturbation theory (MPT) model~\cite{Shirkov:1997wi} is phenomenologically successful. We adopt the MPT model with the preferable parameters $\alpha_{\rm crit}=0.61$ and $\xi=10$ to do our estimation~\footnote{Taking those two values, we obtain a well matching of $\alpha_s$ in both small and large scale regions by using the method of Ref.\cite{Brodsky:2010ur}.}, and by using $\alpha_s(M_{\tau})=0.325\pm0.016$~\cite{Zyla:2020zbs}, we obtain the QCD asymptotic scale under MPT, e.g. $\Lambda_{{\rm QCD},n_f=4}|_{\rm MPT}=236.5^{+27.6}_{-26.5}$ MeV.

\begin{figure}[htb]
\includegraphics[width=0.48\textwidth]{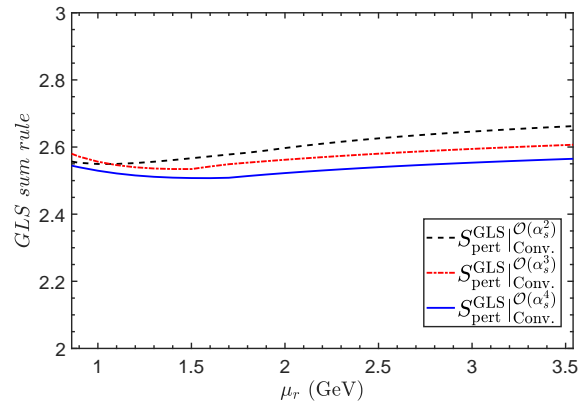}
\caption{The perturbative part of GLS sum rule $S^{\rm GLS}_{\rm pert}(Q^2_0)$ versus the renormalization scale ($\mu_r$) under conventional scale-setting approach up to different perturbative orders.} \label{Convurdepen}
\end{figure}

By setting all input parameters to be their central values, the perturbative part of GLS sum rule $S^{\rm GLS}_{\rm pert}(Q_0^2)$ under conventional scale-setting approach is shown in FIG.~\ref{Convurdepen}. It shows that when more loop terms have been included, $S^{\rm GLS}_{\rm pert}(Q_0^2)$ becomes more steady over different choice of renormalization scale, but there is still sizable scale dependence up to $\mathcal{O}(\alpha_s^4)$ accuracy. Numerically, we have $S^{\rm GLS}_{\rm pert}(Q_0^2)|_{\rm Conv.} =[2.505, 2.562]$ for $\mu_r \in [Q_0/2, 2Q_0]$, and $[2.505, 2.589]$ for $\mu_r \in [Q_0/3, 3Q_0]$; e.g., the net scale errors are $\sim2.3\%$, and $\sim3.3\%$, respectively. We should point out that such small net scale dependence for the prediction up to $\mathcal{O}(\alpha_s^4)$ accuracy is due to the convergent behavior of the perturbative series, e.g., the relative magnitudes of the LO-terms: NLO-terms: N$^2$LO-terms: N$^3$LO-terms $\simeq$ 1: $34\%$: $13\%$: $10\%$ for the case of $\mu_r=Q_0$, and also due to the cancelation of scale dependence among different orders. However the scale error at each perturbative order remains large and unchangeable~\footnote{For an infinite-order pQCD prediction, cancelation among different orders are exact, and there is no need to know the exact value for each order; however for a fixed-order prediction, we need to know their exact values so as to achieve a precise value for the pQCD approximant up to the known orders. An extreme example is from top-pair production, e.g. it has been shown that a correct NLO term is important for explaining the top-pair forward-backward asymmetry at the Tevatron~\cite{Brodsky:2012rj}.}. For example, the $S^{\rm GLS}_{\rm pert}(Q_0^2)$ up to N$^3$LO-level has the following perturbative behavior:
\begin{eqnarray}
S^{\rm GLS}_{\rm pert}(Q_0^2)|_{\rm Conv.} &=& 3(1-0.104^{+0.027}_{-0.008}-0.035^{-0.001}_{+0.002}\nonumber \\
& & \quad -0.014^{-0.006}_{+0.003}-0.010^{-0.003}_{+0.001}) \nonumber \\
&=& 2.511^{+0.051}_{-0.006},
\end{eqnarray}
where the central values are for $\mu_r=Q_0$ GeV, and the errors are obtained by varying $\mu_r \in [Q_0/2, 2Q_0]$. It shows that the absolute scale errors are about $34\%$, $9\%$, $64\%$, and $40\%$ for the LO-terms, NLO-terms, N$^2$LO-terms, and N$^3$LO-terms, respectively.

\begin{figure}[htb]
\includegraphics[width=0.48\textwidth]{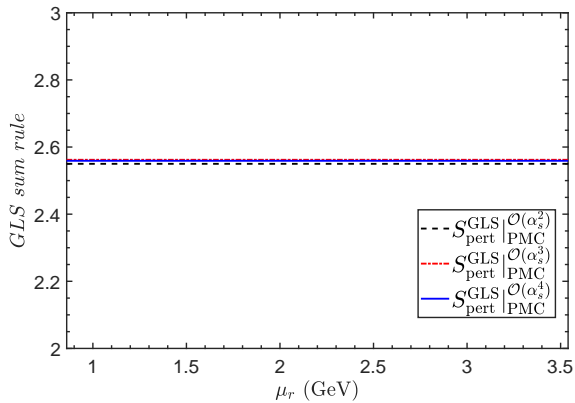}
\caption{The perturbative part of GLS sum rule $S^{\rm GLS}_{\rm pert}(Q_0^2)$ versus the renormalization scale ($\mu_r$) under PMC single-scale approach up to different perturbative orders.} \label{pmcurdepen}
\end{figure}

Similarly, we present $S^{\rm GLS}_{\rm pert}(Q_0^2)$ under PMC single-scale approach in FIG.~\ref{pmcurdepen}. After applying the PMC, the perturbative nature of the GLS sum rule $S^{\rm GLS}_{\rm pert}(Q_0^2)$ is greatly improved: 1) The pQCD convergence $S^{\rm GLS}_{\rm pert}(Q_0^2)$ is improved due to the elimination of divergent renormalon terms, and the relative magnitudes of LO: NLO: N$^2$LO: N$^3$LO of the pQCD series changes to 1: $14\%$: $-7\%$: $-3\%$. This explains why, as shown by FIG.~\ref{pmcurdepen}, the results of $S^{\rm GLS}_{\rm pert}(Q_0^2)$ up to different orders are nearly coincide with each other; 2) There is no renormalization scale ambiguity for PMC prediction, e.g. FIG.~\ref{pmcurdepen} also shows that $S^{\rm GLS}_{\rm pert}(Q_0^2)$ is independent to any choice of $\mu_r$. More explicitly, we have
\begin{eqnarray}
S^{\rm GLS}_{\rm pert}(Q_0^2)|_{\rm PMC}&=& 3(1-0.141-0.020+0.010+0.004) \nonumber \\
&=& 2.559.
\end{eqnarray}
Using Eq.(\ref{scaleQs}), the PMC scale $Q_*$ can be fixed up to N$^2$LL accuracy. For the case of $Q^2=Q_0^2=3$ GeV$^2$, we have
\begin{eqnarray}
\ln\frac{Q^2_*}{Q^2_0}&&=-1.083-1.2141a_s(Q_0)-17.315a^2_s(Q_0), \label{Qstar}
\end{eqnarray}
which leads to $Q_*=0.86$ GeV. $Q_*$ is independent to the choice of $\mu_r$. Due to its perturbative nature, as a conservative estimation of the unknown perturbative terms, we take the magnitude of the last known term as the unknown N$^3$LL term, which leads to a scale shift of $\Delta Q_* \simeq \left(^{+0.09}_{-0.08}\right)$ GeV and then
\begin{equation}
\Delta S^{\rm GLS}_{\rm pert}(Q_0^2)|_{\rm  PMC}=\pm0.015.
\end{equation}
This error is called as {\it the first kind of residual scale dependence due to unknown higher-order terms}~\cite{Zheng:2013uja}.

\begin{table}[htb]
\begin{tabular}{ccc}
\hline

      \raisebox {0ex}[0pt]
            & ~~$\rm N^3LO$~~    &  ~~$\rm N^4LO$~~  \\
\hline
      \raisebox {0ex}[0pt]{EC}
         & $-0.010^{-0.003}_{+0.001}$ & - \\
\hline
      \raisebox {0ex}[0pt]{PAA}
         & ~~[1/1]:$-0.005^{-0.005}_{+0.002}$~~ & ~~[2/1]:$-0.007^{+0.000}_{-0.001}$~~ \\
\hline
\end{tabular}
\caption{The preferable diagonal-type PAA predictions of the $\rm N^3LO$, and $\rm N^4LO$ terms of $a_{F_3}|_{\rm Conv.}(Q_0)$ under conventional scale-setting approach. The central value is for $\mu_r=Q_0$, and the errors are for $\mu_r\in[Q_0/2, 2Q_0]$. ``EC" is the exact result using the known series. }  \label{convpaa}
\end{table}

\begin{table}[htb]
\centering
\begin{tabular}{ccc}
\hline
      \raisebox {0ex}[0pt]{~~~~}
              & ~~$\rm N^3LO$~~    & ~~$\rm N^4LO$~~  \\
\hline
{~~EC~~}    & ~~$0.004$~~ & ~~-~~ \\
\hline
{~~PAA~~}   & ~~[0/2]:$0.003$~~ & ~~[0/3]:$-0.0001$~~ \\
\hline
\end{tabular}
\caption{The preferable [0/$n$-1]-type PAA predictions of the, $\rm N^3LO$, and $\rm N^4LO$ terms of $a_{F_3}|_{\rm PMC}(Q_0)$ under the PMC scale-setting approach, which is independent of any choice of $\mu_r$. ``EC" is the exact prediction using the known series. }  \label{pmcpaa}
\end{table}

For a perturbative prediction, it is helpful to give a reliable prediction of the magnitude of unknown perturbative terms. The Pad\'{e} approximation approach (PAA)~\cite{Basdevant:1972fe, Samuel:1992qg, Samuel:1995jc} is one of such approaches for estimating the $(n+1)_{\rm th}$-order coefficient in a given $n_{\rm th}$-order perturbative series, which give feasible conjectures on the higher-order behavior of the series. For a known pQCD series $\rho(Q)=\sum^n_{i=1} C_i a_s^i$, its PAA $[N/M]$-type fractional generating function is defined as
\begin{eqnarray}
\rho^{N/M}(Q)&=&a_s\times\frac{b_0+b_1 a_s+\cdots+b_N a_s^N}{1+c_1 a_s+\cdots+c_M a_s^M} \nonumber \\
&=&\sum^n_{i=1} C_i a_s^i+C_{n+1} a_s^{n+1}+\cdots.
\end{eqnarray}
where $N+M+1=n$ ($M\geq1$). The unknown $(n+1)_{\rm th}$-order coefficient $C_{n+1}$ can be expressed by the known $C_{1,\cdots,n}$ by expanding the fractional generating function over $a_s$, e.g. we first express all the coefficients $\{b_0,\cdots,b_N\}$ and $\{c_1,\cdots,c_M\}$ by the known coefficients $C_{1,\cdots,n}$, and then get the coefficient $C_{n+1}$ over $\{b_i\}$ and $\{c_i\}$, which can be finally expressed by $\{C_i\}$. For the present case, we have $\rho(Q)=a_{F_3}|_{\rm PMC}$ or $a_{F_3}|_{\rm Conv.}$, and $n=4$. The effectiveness of the PAA depends heavily on how well we know the perturbative series, such as the number of loop terms, the perturbative convergence, the accuracy of each loop terms, etc..

Following the standard procedures described in detail in Ref.\cite{Du:2018dma}, we give the PAA predictions of $\rm N^3LO$, and $\rm N^4LO$ terms in Table~\ref{convpaa} and Table~\ref{pmcpaa}. In Table~\ref{convpaa}, we give the preferable diagonal-type PAA prediction for conventional series~\cite{Gardi:1996iq}; and in Table~\ref{pmcpaa}, we give the preferable [0/$n$-1]-type PAA predictions for PMC series~\cite{Du:2018dma}. Because of large scale dependence for each loop-terms, the PAA prediction based on conventional series also has large scale dependence: if taking $\mu_r\in[Q_0/2, 2Q_0]$, the PAA predicted N$^3$LO-term is about $33\%$-$77\%$ of the exact N$^3$LO-term~\footnote{There is another diagonal-type generating function for the N$^3$LO conventional series, e.g. the [1/2]-type, whose predicted N$^4$LO-term is $-0.016^{+0.007}_{-0.019}$, which however could be excluded due to its pure convergence and much larger scale dependence.}. On the other hand, the PMC conformal series, which is scheme-and-scale independent at any loop terms, provides a helpful basis for a reliable prediction of the unknown higher-orders. Its predicted N$^3$LO-term and N$^4$LO-term are exactly scale invariant, and the N$^3$LO-term is about $75\%$ of the exact N$^3$LO-term, and the predicted N$^4$LO is negligible, which is only $\sim 3\%$ of the N$^3$LO-term, indicating the present known ${\cal O}(\alpha_s^4)$ PMC series is already at very high accuracy. This error could be called as {\it the second kind of residual scale dependence due to unknown higher-order terms}. Then by using Eq.(\ref{eq:gls}), our prediction of the magnitude of the N$^4$LO-terms of the GLS sum rule is
\begin{eqnarray}
\Delta S^{\rm GLS}(Q_0^2)|^{\rm N^4LO}_{\rm Conv.} &=& \pm 0.021, \\
\Delta S^{\rm GLS}(Q_0^2)|^{\rm N^4LO}_{\rm PMC}   &=& \pm 0.0003.
\end{eqnarray}

The squared average of the above two residual scale dependence leads to a net perturbative error due to uncalculated higher-order terms under conventional and PMC scale-setting approaches, i.e.
\begin{eqnarray}
\Delta S^{\rm GLS}(Q_0^2)|^{\text{High order}}_{\rm  Conv.} &\simeq& \left(^{+0.055}_{-0.022}\right), \\
\Delta S^{\rm GLS}(Q_0^2)|^{\text{High order}}_{\rm  PMC}   &\simeq& \pm 0.015.
\end{eqnarray}

There is theoretical error due to $\Delta\alpha_s(M_Z)$. As an estimation, using $\alpha_s(M_Z)=0.1179\pm0.0010$~\cite{Zyla:2020zbs}, we obtain $\Lambda_{{\rm QCD},n_f=4}=291.7^{+14.9}_{-14.4}$ MeV, which lead to
\begin{eqnarray}
\Delta S^{\rm GLS}(Q_0^2)|^{\Delta\alpha_s(M_Z)}_{\rm Conv.} &=& \left(^{+0.019}_{-0.022}\right), \\
\Delta S^{\rm GLS}(Q_0^2)|^{\Delta\alpha_s(M_Z)}_{\rm PMC}   &=& \left(^{+0.018}_{-0.019}\right).
\end{eqnarray}
This shows that for the present four-loop analysis, the accuracy of the reference point $\alpha_s(M_Z)$ is of the same importance as the errors caused by the unknown perturbative terms.

Our final prediction for $S^{\rm GLS}(Q_0^2)$ using known N$^3$LO perturbative series is
\begin{eqnarray}
S^{\rm GLS}(Q_0^2)|_{\rm Conv.}&=& 2.511^{+0.058}_{-0.031} \label{result1}
\end{eqnarray}
and
\begin{eqnarray}
S^{\rm GLS}(Q_0^2)|_{\rm PMC}  &=& 2.559^{+0.023}_{-0.024}. \label{result2}
\end{eqnarray}
where the errors are mean square of $\Delta S^{\rm GLS}(Q_0^2)|^{\text{High order}}$ and $\Delta S^{\rm GLS}(Q_0^2)|^{\Delta\alpha_s(M_Z)}$. The conventional and PMC predictions are consistent with each other within errors. The perturbative error of the PMC single-scale approach is independent to any choice of $\mu_r$, while the perturbative error for conventional scale-setting approach is for $\mu_r \in[Q_0/2, 2Q_0]$. The accuracy of the pQCD prediction depends strongly on the exact value of $\alpha_s$, e.g. the precise magnitude of $\alpha_s$ and the precise value of the reference point $\alpha_s(M_Z)$. After applying the PMC, the correct effective $\alpha_s$ can be achieved by applying the RGE, its resultant more convergent pQCD series also leads to a much smaller residual scale dependence, thus a more reliable and precise pQCD prediction without renormalization scale ambiguity can be achieved.

\begin{figure}[htb]
\includegraphics[width=0.48\textwidth]{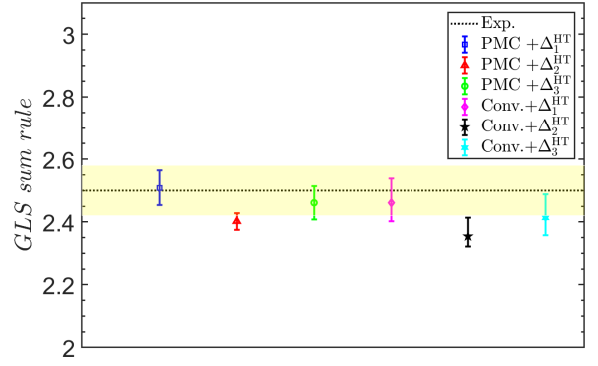}
\caption{The GLS sum rules $S^{\rm GLS}(Q_0^2)$ in comparison to the experimental result. The dotted line is the CCFR measured value $S^{\rm GLS}_{\rm exp}(Q_0^2)=2.50\pm0.08$~\cite{Leung:1992yx} and the shaded band shows its uncertainty. In theoretical predictions, both the perturbative contributions under conventional and PMC single-scale setting approaches and three typical non-perturbative high-twist contributions $\Delta^{\rm HT}_{1,2,3}$ have been included. }
\label{GLSerror}
\end{figure}

By adding the nonperturbative high-twist contribution $\Delta^{\rm HT}$ to the GLS sum rule, one can do a comparison with its measured value. Using three typical $\Delta^{\rm HT}$ as listed in the beginning of the paper, we present such a comparison with the measured value of $S^{\rm GLS}(Q_0^2)$ in Fig.~\ref{GLSerror}, whose error is the mean square of the systematic and statistical errors. Under the same choice of $\Delta^{\rm HT}$, Fig.~\ref{GLSerror} shows the PMC predictions are in better agreement with the data in comparison to those of conventional ones.

\begin{figure}[htb]
\includegraphics[width=0.48\textwidth]{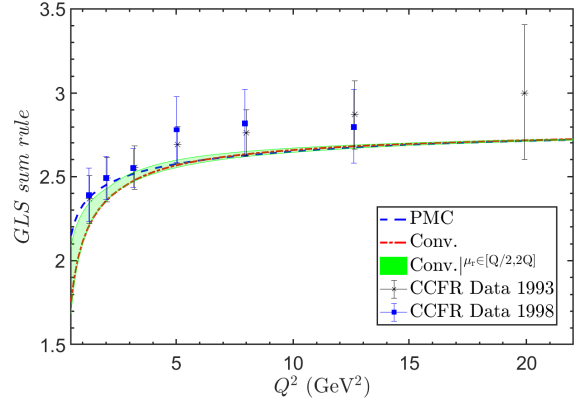}
\caption{The GLS sum rules $S^{\rm GLS}(Q^2)$ in small $Q^2$-region under conventional and PMC single-scale approaches. The experimental data given by the CCFR collaborations~\cite{Leung:1992yx, Kim:1998kia} are presented as a comparison. The dotted line is the conventional pQCD prediction for $\mu_r=Q$ GeV and the shaded band is for $\mu_r \in [Q/2, 2Q]$. The dashed line is the scale-invariant PMC prediction. $\Delta^{\rm HT}=-0.15$. }
\label{GLS1}
\end{figure}

\begin{figure}[htb]
\includegraphics[width=0.48\textwidth]{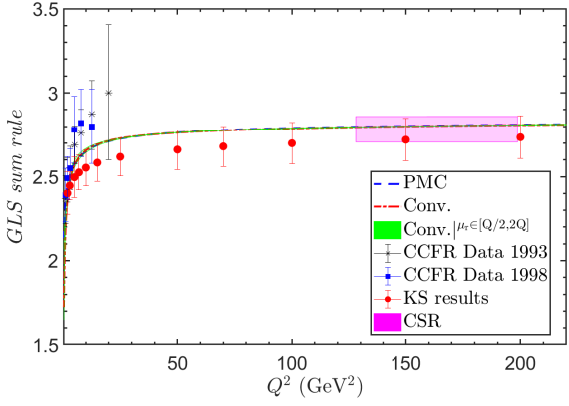}
\caption{The GLS sum rule $S^{\rm GLS}(Q^2)$ in large $Q^2$-region under conventional and PMC single-scale approaches. As a comparison, the experimental data given by the CCFR collaborations~\cite{Leung:1992yx, Kim:1998kia}, and the extrapolation of experimental data given by Ref.\cite{Huang:2020gic} (labeled as ``CSR"), Ref.\cite{Kataev:1994rj} (Labeled as ``KS results") have also been presented. }
\label{GLS2}
\end{figure}

The above treatments for the case of $Q^2=Q^2_0$ is adaptable for any $Q^2$ value. We present the $Q^2$-dependence of GLS sum rule under conventional and PMC scale-setting approaches in FIGs.~(\ref{GLS1}, \ref{GLS2}). FIG.~\ref{GLS1} shows its behavior in small $Q^2$-region and FIG.~\ref{GLS2} shows its behavior in large $Q^2$-region. The shaded bands show the scale uncertainties for conventional scale-setting approach. As shown by FIG.~\ref{GLSerror}, the case of $\Delta^{\rm HT}=\Delta^{\rm HT}_{1}$ shows better agreement with the data, so we have implicitly taken $\Delta^{\rm HT}=-0.15$. In large $Q^2$-region, the power suppressed $\Delta^{\rm HT}$ gives negligible contribution and the GLS sum rule is dominated by its perturbative part. Our prediction is consistent with $S^{\rm GLS}|_{\rm CSR}(Q^{2} =158^{+39}_{-30} {\rm GeV}^2) = 2.781^{+0.078}_{-0.075}$~\cite{Huang:2020gic}, which is derived by using the scheme-independent commensurate scale relation. And in large $Q^2$-region, the GLS sum rule agrees with the extrapolation of experimental data given by Ref.\cite{Kataev:1994rj}. The conventional scale dependence becomes small when $Q^2$ becomes large, and when $Q^2$ is large enough, the conventional prediction almost coincides with the scale-invariant PMC prediction. This agrees with the conventional wisdom that when we have known enough higher-order terms for a pQCD series, the pQCD prediction becomes scale invariant.

To make the goodness-of-fit more clearly, we calculate the reduced $\chi^2$ for $S^{\rm GLS}(Q^2)$, e.g. $\chi^2/N$ ($N$ stands for the degrees of freedom), which is defined as
\begin{eqnarray}
\chi^2/N=\frac{1}{N}\sum^N_{i=0}\left[\frac{y^{\rm Exp.}_{i} - y^{\rm The.}(x_i)}{\sigma_i}\right]^2,
\end{eqnarray}
where $y^{\rm Exp.}$ is the measured value at each point, $\sigma_i$ is its uncertainty, $y^{\rm The.}$ is the theoretical prediction by setting all input parameters to be their central values. Thirteen data~\cite{Leung:1992yx, Kim:1998kia} are adopted, i.e. $N=13$. We obtain $\chi^2/N=0.46$ for PMC single-scale approach and $\chi^2/N=0.83$ for conventional scale-setting approach, indicating both predictions under conventional scale-setting and PMC single-scale approaches are highly consistent with the data. Due to a smaller $\chi^2/N$, which corresponds to the $p$-value $\sim 90\%-95\%$~\cite{Zyla:2020zbs}, the PMC prediction represents a better fit of the experiment data.

As an addendum, one can inversely obtain a precise prediction on the non-perturbative high-twist contribution $\Delta^{\rm HT}$ by comparing the experimental data with the highly precise PMC prediction up to ${\cal O}(\alpha_s^4)$-level. Using the CCFR experimental data~\cite{Leung:1992yx, Kim:1998kia} for GLS sum rule, we obtain $\Delta^{\rm HT}=-0.061^{+0.031}_{-0.039}$ (the error is caused by $\Delta\alpha_s(M_Z)=\pm0.0010$), which corresponds to the minimum value of $\chi^2/N|_{\rm min}\simeq 0.40$.

As a summary, we have presented a detailed analysis for GLS sum rule $S^{\rm GLS}(Q^2)$ by applying the PMC single-scale approach. It is essential to finish as more perturbative terms as possible. The conventional scale-setting approach to fixed-order pQCD predictions is based on an arbitrary choice of the renormalization scale, together with an arbitrary range. And it is found that by further applying the PMC single-scale approach to the conventional pQCD series, a renormalization scheme-and-scale invariant fixed-order pQCD prediction can be obtained, which also shows better agreement with the data. The residual scale dependence due to unknown perturbative terms can be greatly suppressed by applying the PMC to the present known ${\cal O}(\alpha_s^4)$-level series. Therefore, we think that after finishing the usual perturbative calculations, the PMC single-scale approach is an important step forward for achieving precise pQCD predictions.

\hspace{2cm}

\noindent {\bf Acknowledgments:} This work was supported in part by the Natural Science Foundation of China under Grant No.11625520 and No.12047564, by the Fundamental Research Funds for the Central Universities under Grant No.2020CQJQY-Z003, and by the Chongqing Graduate Research and Innovation Foundation under Grant No.ydstd1912.

\appendix

\section*{Appendix: The reduced coefficients ${\hat r}_{i,j}$ and the coefficients $T_{i}$}

Up to ${\cal O}(\alpha_s^4)$-order level, the reduced coefficients ${\hat r}_{i,j}$ for the perturbative series of GLS sum rule are
\begin{eqnarray}
{\hat r}_{1,0} &=& \frac{3}{4}{\gamma^{\rm ns}_1}, \nonumber \\
{\hat r}_{2,0} &=& \frac{3}{4}{\gamma^{\rm ns}_2}-\frac{9}{16}\big({\gamma^{\rm ns}_1}\big)^2, \nonumber \\
{\hat r}_{3,0} &=& \frac{3}{4}({\gamma^{\rm ns}_3}+{\gamma^{\rm s}_3 n_f})-\frac{9}{8}{\gamma^{\rm ns}_2}{\gamma^{\rm ns}_1}+\frac{27}{64}\big({\gamma^{\rm ns}_1}\big)^3,\nonumber  \\
{\hat r}_{4,0} &=& \frac{3}{4}({\gamma^{\rm ns}_4}+{\gamma^{\rm s}_4 n_f})-\frac{9}{8}({\gamma^{\rm ns}_3}+{\gamma^{\rm s}_3 n_f}){\gamma^{\rm ns}_1}-\frac{9}{16}\big({\gamma^{\rm ns}_2}\big)^2\nonumber \\
&&+\frac{81}{64} {\gamma^{\rm ns}_2} \big({\gamma^{\rm ns}_1}\big)^2-\frac{81}{256}\big({\gamma^{\rm ns}_1}\big)^4, \nonumber \\
{\hat r}_{2,1} &=& {3 \over 4}{\Pi^{\rm ns}_1}+{K_1},  \nonumber\\
{\hat r}_{3,1} &=& {3 \over 4}{\Pi^{\rm ns}_2}+{1 \over 2}{K_2}-\frac{\gamma^{\rm ns}_1}{4}\left(\frac{3}{2}{K_1}+{9 \over 4}{\Pi^{\rm ns}_1}\right),\;\; {\hat r}_{3,2}=0,  \nonumber\\
{\hat r}_{4,1} &=& {3 \over 4}({\Pi^{\rm ns}_3}+{\Pi^{\rm s}_3 n_f})+{1 \over 3}{K_3}-{1 \over 4}{\gamma^{\rm ns}_1}\left({K_2}+3{\Pi^{\rm ns}_2}\right)\nonumber \\
&&-\frac{\gamma^{\rm ns}_2}{4}\left({K_1}+{3 \over 2}{\Pi^{\rm ns}_1}\right) +\frac{\big({\gamma^{\rm ns}_1}\big)^2}{16} \left(3{K_1}+\frac{27}{4}{\Pi^{\rm ns}_1}\right),  \nonumber\\
{\hat r}_{4,2} &=& -\frac{3}{16}\big({\Pi^{\rm ns}_1}\big)^2-{1 \over 4}{K_1} {\Pi^{\rm ns}_1}, \;\; {\hat r}_{4,3} = 0, \nonumber
\end{eqnarray}
where the expressions for $\gamma^{\rm ns}_i$, $\Pi^{\rm ns}_i$, $\gamma^{\rm s}_i$, $\Pi^{\rm s}_i$, and $K_i$ can be found in Refs.\cite{Baikov:2010je, Baikov:2012zn, Baikov:2012zm}. The PMC scale $Q_*$ can be determined up to N$^2$LL accuracy, whose coefficients are
\begin{eqnarray}
T_0=&&-\frac{{\hat r}_{2,1}}{{\hat r}_{1,0}}, \label{Tij1} \\
T_1=&&\frac{ \beta _0 ({\hat r}_{2,1}^2-{\hat r}_{1,0} {\hat r}_{3,2})}{{\hat r}_{1,0}^2}+\frac{2 ({\hat r}_{2,0} {\hat r}_{2,1}-{\hat r}_{1,0} {\hat r}_{3,1})}{{\hat r}_{1,0}^2} \label{Tij2}
\end{eqnarray}
and
\begin{eqnarray}
T_2=&&\frac{3 \beta _1 ({\hat r}_{2,1}^2-{\hat r}_{1,0} {\hat r}_{3,2})}{2 {\hat r}_{1,0}^2} \nonumber\\
&&+\frac{4({\hat r}_{1,0} {\hat r}_{2,0} {\hat r}_{3,1}-{\hat r}_{2,0}^2 {\hat r}_{2,1})+3({\hat r}_{1,0} {\hat r}_{2,1} {\hat r}_{3,0}-{\hat r}_{1,0}^2 {\hat r}_{4,1})}{ {\hat r}_{1,0}^3} \nonumber \\
&&+\frac{ \beta _0  (4 {\hat r}_{2,1} {\hat r}_{3,1} {\hat r}_{1,0}-3 {\hat r}_{4,2} {\hat r}_{1,0}^2+2 {\hat r}_{2,0} {\hat r}_{3,2} {\hat r}_{1,0}-3 {\hat r}_{2,0} {\hat r}_{2,1}^2)}{ {\hat r}_{1,0}^3}\nonumber\\
&&+\frac{ \beta _0^2 (2 {\hat r}_{1,0} {\hat r}_{3,2} {\hat r}_{2,1}- {\hat r}_{2,1}^3- {\hat r}_{1,0}^2 {\hat r}_{4,3})}{ {\hat r}_{1,0}^3}.
\end{eqnarray}


\begin{thebibliography}{00}

\bibitem{Gross:1969jf}
  D.~J.~Gross and C.~H.~Llewellyn Smith,
  Nucl.\ Phys.\ B {\bf 14}, 337 (1969).

\bibitem{Leung:1992yx}
  W.~C.~Leung {\it et al.},
  Phys.\ Lett.\ B {\bf 317}, 655 (1993).

\bibitem{Kim:1998kia}
  J.~H.~Kim {\it et al.},
  Phys.\ Rev.\ Lett.\  {\bf 81}, 3595 (1998).

\bibitem{Fajfer:1985fw}
  S.~Fajfer and R.~J.~Oakes,
  Phys.\ Lett.\ B {\bf 163}, 385 (1985).

\bibitem{Ross:1993gb}
  G.~G.~Ross and R.~G.~Roberts,
  Phys.\ Lett.\ B {\bf 322}, 425 (1994).

\bibitem{Braun:1986ty}
  V.~M.~Braun and A.~V.~Kolesnichenko,
  Nucl.\ Phys.\ B {\bf 283}, 723 (1987).

\bibitem{Anselmino:1995re}
  M.~Anselmino, F.~Caruso and E.~Levin,
  Phys.\ Lett.\ B {\bf 358}, 109 (1995).

\bibitem{Dasgupta:1996hh}
  M.~Dasgupta and B.~R.~Webber,
  Phys.\ Lett.\ B {\bf 382}, 273 (1996).

\bibitem{Kataev:1999bp}
  A.~L.~Kataev, G.~Parente and A.~V.~Sidorov,
  Nucl.\ Phys.\ B {\bf 573}, 405 (2000).

\bibitem{Nath:2016phi}
  N.~M.~Nath, A.~Mukharjee, M.~K.~Das and J.~K.~Sarma,
  Commun.\ Theor.\ Phys.\  {\bf 66}, 663 (2016).

\bibitem{Zyla:2020zbs}
  P.~A.~Zyla {\it et al.} [Particle Data Group],
  PTEP {\bf 2020}, 083C01 (2020).

\bibitem{Gorishnii:1985xm}
  S.~G.~Gorishnii and S.~A.~Larin,
  Phys.\ Lett.\ B {\bf 172}, 109 (1986).

\bibitem{Zijlstra:1993sh}
  E.~B.~Zijlstra and W.~L.~van Neerven,
  Nucl.\ Phys.\ B {\bf 417}, 61 (1994).

\bibitem{Larin:1991tj}
  S.~A.~Larin and J.~A.~M.~Vermaseren,
  Phys.\ Lett.\ B {\bf 259}, 345 (1991).

\bibitem{Chyla:1992cg}
  J.~Chyla and A.~L.~Kataev,
  Phys.\ Lett.\ B {\bf 297}, 385 (1992).

\bibitem{Hinchliffe:1996hc}
  I.~Hinchliffe and A.~Kwiatkowski,
  Ann.\ Rev.\ Nucl.\ Part.\ Sci.\  {\bf 46}, 609 (1996).

\bibitem{Nath:2015sxi}
  N.~M.~Nath, M.~K.~Das and J.~K.~Sarma,
  Indian J.\ Phys.\  {\bf 90}, 117 (2016).

\bibitem{Baikov:2010je}
  P.~A.~Baikov, K.~G.~Chetyrkin, and J.~H.~Kuhn,
  Phys.\ Rev.\ Lett.\  {\bf 104}, 132004 (2010).

\bibitem{Baikov:2012zn}
  P.~A.~Baikov, K.~G.~Chetyrkin, J.~H.~Kuhn, and J.~Rittinger,
  Phys.\ Lett.\ B {\bf 714}, 62 (2012).

\bibitem{Wu:2013ei}
  X.~G.~Wu, S.~J.~Brodsky, and M.~Mojaza,
  Prog.\ Part.\ Nucl.\ Phys.\  {\bf 72}, 44 (2013).

\bibitem{Wu:2014iba}
  X.~G.~Wu, Y.~Ma, S.~Q.~Wang, H.~B.~Fu, H.~H.~Ma, S.~J.~Brodsky, and M.~Mojaza,
  Rep.\ Prog.\ Phys.\  {\bf 78}, 126201 (2015).

\bibitem{Wu:2019mky}
  X.~G.~Wu, J.~M.~Shen, B.~L.~Du, X.~D.~Huang, S.~Q.~Wang, and S.~J.~Brodsky,
  Prog.\ Part.\ Nucl.\ Phys.\ {\bf 108}, 103706 (2019).

\bibitem{Brodsky:2011ta}
  S.~J.~Brodsky and X.~G.~Wu,
  Phys.\ Rev.\ D {\bf 85}, 034038 (2012).

\bibitem{Brodsky:2012rj}
  S.~J.~Brodsky and X.~G.~Wu,
  Phys.\ Rev.\ Lett.\ {\bf 109}, 042002 (2012).

\bibitem{Mojaza:2012mf}
  M.~Mojaza, S.~J.~Brodsky, and X.~G.~Wu,
  Phys.\ Rev.\ Lett.\  {\bf 110}, 192001 (2013).

\bibitem{Brodsky:2013vpa}
  S.~J.~Brodsky, M.~Mojaza, and X.~G.~Wu,
  Phys.\ Rev.\ D {\bf 89}, 014027 (2014).

\bibitem{Shen:2017pdu}
  J.~M.~Shen, X.~G.~Wu, B.~L.~Du, and S.~J.~Brodsky,
  Phys.\ Rev.\ D {\bf 95}, 094006 (2017).

\bibitem{Wu:2018cmb}
  X.~G.~Wu, J.~M.~Shen, B.~L.~Du, and S.~J.~Brodsky,
  Phys.\ Rev.\ D {\bf 97}, 094030 (2018).

\bibitem{Zheng:2013uja}
  X.~C.~Zheng, X.~G.~Wu, S.~Q.~Wang, J.~M.~Shen, and Q.~L.~Zhang,
  J. High Energy Phys.  {\bf 10}, 117 (2013).


\bibitem{Baikov:2008jh}
  P.~A.~Baikov, K.~G.~Chetyrkin, and J.~H.~Kuhn,
  Phys.\ Rev.\ Lett.\  {\bf 101}, 012002 (2008).

\bibitem{Shirkov:2012ux}
  D.~V.~Shirkov,
  Phys.\ Part.\ Nucl.\ Lett.\  {\bf 10}, 186 (2013).

\bibitem{Shirkov:1997wi}
  D.~V.~Shirkov and I.~L.~Solovtsov,
  Phys.\ Rev.\ Lett.\  {\bf 79}, 1209 (1997).

\bibitem{Brodsky:2010ur}
  S.~J.~Brodsky, G.~F.~de Teramond, and A.~Deur,
  Phys.\ Rev.\ D {\bf 81}, 096010 (2010).

\bibitem{Webber:1998um}
  B.~R.~Webber,
  JHEP {\bf 9810}, 012 (1998).

\bibitem{Badalian:2001by}
  A.~M.~Badalian and D.~S.~Kuzmenko,
  Phys.\ Rev.\ D {\bf 65}, 016004 (2001).

\bibitem{Cornwall:1981zr}
  J.~M.~Cornwall,
  Phys.\ Rev.\ D {\bf 26}, 1453 (1982).

\bibitem{Godfrey:1985xj}
  S.~Godfrey and N.~Isgur,
  Phys.\ Rev.\ D {\bf 32}, 189 (1985).

\bibitem{Zhang:2014qqa}
  Q.~L.~Zhang, X.~G.~Wu, X.~C.~Zheng, S.~Q.~Wang, H.~B.~Fu, and Z.~Y.~Fang,
  Chin.\ Phys.\ Lett.\  {\bf 31}, 051202 (2014).

\bibitem{Basdevant:1972fe}
  J.~L.~Basdevant,
  Fortsch.\ Phys.\ {\bf 20}, 283 (1972).

\bibitem{Samuel:1992qg}
  M.~A.~Samuel, G.~Li, and E.~Steinfelds,
  Phys.\ Lett.\ B {\bf 323}, 188 (1994).

\bibitem{Samuel:1995jc}
  M.~A.~Samuel, J.~R.~Ellis, and M.~Karliner,
  Phys.\ Rev.\ Lett.\  {\bf 74}, 4380 (1995).

\bibitem{Du:2018dma}
  B.~L.~Du, X.~G.~Wu, J.~M.~Shen, and S.~J.~Brodsky,
  Eur.\ Phys.\ J.\ C {\bf 79}, 182 (2019).

\bibitem{Gardi:1996iq}
  E.~Gardi,
  Phys.\ Rev.\ D {\bf 56}, 68 (1997).

\bibitem{Huang:2020gic}
  X.~D.~Huang, X.~G.~Wu, Q.~Yu, X.~C.~Zheng, J.~Zeng and J.~M.~Shen,
  arXiv:2010.08910 [hep-ph].

\bibitem{Kataev:1994rj}
  A.~L.~Kataev and A.~V.~Sidorov,
  Phys.\ Lett.\ B {\bf 331}, 179 (1994).

\bibitem{Baikov:2012zm}
  P.~A.~Baikov, K.~G.~Chetyrkin, J.~H.~Kuhn, and J.~Rittinger,
  JHEP {\bf 1207}, 017 (2012).

\end{thebibliography}
\end{document}